\documentstyle[epsfig,12pt]{article}
\begin{document}
\author{ Andr\'e Roug\'e\thanks{Andre.Rouge@in2p3.fr}\\
 LPNHE Ecole Polytechnique-IN2P3/CNRS\\
 F-91128 Palaiseau Cedex}
\date{October 1, 2000}
\title{\mbox{~}\\Tau lepton Michel parameters and new physics}
\maketitle\mbox{~}\\
\begin{abstract}
A complete discussion of the constraints on the Michel parameters and
the ambiguities of their interpretation is presented. Estimators of
new physics, optimized for a very wide class of hypotheses and models
are proposed.
\end{abstract}
\mbox{~}\\[-13cm]
\begin{flushright}
X-LPNHE 00/01
\end{flushright}
\mbox{~}\\[11.25cm]

\section{Introduction}
\label{cha:intro}

The leptonic decay of a charged lepton $l_1^-\to l_2^-\,\nu_1\,\bar\nu_2$
is described by its Michel parameters~[1-4]
Complementary parameters must be
introduced~\cite{cj} to describe the secondary lepton polarization.
The measurements of all the parameters and  the
cross-section for the $\nu_1\,l_2^-\to l_1^-\,\nu_2$ reaction allow
to put upper limits on all the non-V-A amplitudes~\cite{cj,fetscha}
and ``prove experimentally'' the V-A structure of the weak interaction.

In the case of the   $\mu$ lepton, this program was followed and
successfully completed fifteen years ago~\cite{fetscha,pdg}.

For the $\tau$ lepton, it is customary~[7-9]
 to present the experimental
results, in a similar way, as upper values of the possible coupling constants.
However, the validity of the Standard Model is today so well
established, that the interest of  $\tau$ Michel parameter
measurements is no more to prove the V-A structure but to look
for small deviations from its predictions that would be evidence
of new physics. Moreover, the additional measurements, which are
necessary to  complete the experimental proof of the V-A structure, will not be performed in
a foreseeable future, while large statistics of $\tau$ pairs will soon
be available at B-factories, allowing more precise measurements of the 
Michel parameters.  

Therefore it seems worthwhile to look for  the com\-binations of the
measured parameters which are
the most sensitive to non-standard effects. This is the aim of the
present paper.

\section{The general framework}
\label{sec:lorstru}
If only the momentum of the final state charged lepton is  measured,
the decay of a polarized $\tau$ 
is entirely described, in its centre-of-mass, by the
distribution
\begin{equation}\label{eq:decgencm}
\frac{1}{x^2\Gamma}\frac{d\Gamma}{d\Omega dx}=W_0(x)+P_\tau\, W_1(x)\,\cos\theta~,
\end{equation}
where $P_\tau$ is the $\tau$ polarization, $\theta$ the angle between the
polarization  and the charged lepton momentum and $x=E_l/E_l^{\mathrm{max}}$ the
normalized lepton energy.\\

Taking advantage of the weak interaction short range, the decay can be
 represented
by the most general four fermion contact interaction~[1-4],
 written below in the helicity 
projection formalism~[6,10-12]:
 \begin{eqnarray}
  \label{e:matrix}
\lefteqn{    {\cal M} = 4\frac{G_{\tau l}}{\sqrt{2}} 
            \sum_{{\scriptstyle \gamma=S,V,T \atop\scriptstyle i,j=R,L}} g^{\gamma}_{ij}
            \,\langle \bar{l}_i\vert\Gamma^\gamma\vert\left(\nu_l\right)_n\rangle
            \langle\left(\bar{\nu}_{\tau}\right)_m\vert\Gamma_{\gamma}\vert\tau_j\rangle, 
            }\hspace{6cm}\\[-3ex]\nonumber&& l=e,\mu~.
\end{eqnarray}
 $G_{\tau l}$ is the absolute coupling strength; the $g^{\gamma}_{ij}$ are ten complex 
coupling constants describing the relative contribution of scalar ($\Gamma^{S} = 1$), 
vector ($\Gamma^{V} = \gamma^{\mu}$), and tensor 
($\Gamma^{T} =\frac{1}{\sqrt{2}}\sigma^{\mu\nu}$)
interactions, respectively, for given chiralities $j,i$ of the $\tau$ and the charged decay
lepton. The neutrino chiralities $n$ and $m$ are  uniquely defined for a given set 
\{$\gamma,i,j$\} by the chirality selections rules: conservation for vector
coupling, reversal for scalar and tensor.

\begin{table}[hbt]
\caption{\label{tab:neutral}\protect\small 
Contributions of neutral currents to the $g^\gamma_{ij}$ coupling constants}
\begin{center}
\begin{tabular}{cl}\hline
$
\begin{array}{c}
\hbox{Neutral}\\
\hbox{current}
\end{array}
$
&\multicolumn{1}{c}
{Charged coupling constants}
\\\hline
V&\rule{0ex}{2.5ex}$
\begin{array}{r}
g^V_{LL}\hspace{1em}g^V_{RR}\hspace{2em}
g^S_{LL}\hspace{2.5em}g^S_{RR}
\end{array}
$
\rule[-2.5ex]{0ex}{1ex}\\
S&
$
\begin{array}{r}g^V_{LR} 
\hspace{1em}g^V_{RL}\\\rule{0em}{1ex}\end{array}
\hspace{1em}
\begin{array}{r}
 g^S_{LR}\hspace{3em}g^T_{LR}\\[1ex]
 g^S_{LR}=\hspace{1em}2g^T_{LR}\end{array}
\hspace{1em}
\begin{array}{r}
g^S_{RL}\hspace{3em}g^T_{RL}\\[1ex]
g^S_{RL}=\hspace{1em}2g^T_{RL}
\end{array} 
$
\rule[-4ex]{0ex}{1ex}\\\rule{0ex}{5ex}
T&\hspace{6.2em}$\begin{array}{r}
g^S_{LR}\hspace{2.5em}g^T_{LR}\\[1ex]
g^S_{LR}=-6g^T_{LR}
    \end{array}
\hspace{1.3em}
    \begin{array}{r}
g^S_{RL}\hspace{2.5em}g^T_{RL}\\[1ex]
g^S_{RL}=-6g^T_{RL}
     \end{array}$\rule[-3.5ex]{0ex}{1ex}
\\\hline
\end{tabular}\\[-.5cm]
\end{center}
\end{table}

The same matrix element describes also the contributions of possible
lepton-number-violating neutral currents~\cite{ll}. The relationship
between the neutral currents and the   $g^{\gamma}_{ij}$ constants,
as given by the Fierz transformation, is summarized in Table~\ref{tab:neutral}.\\

 The matrix element (\ref{e:matrix}) can be used to compute the
 functions $W_0$ and $W_1$ 
of Eq.~\ref{eq:decgencm}. 
They are described 
by the four real Michel parameters:
$\rho$ and  $\eta$ for $W_0$, $\xi$ and $\xi\delta$ for $W_1$.

\begin{figure}[ht]
\centerline{\epsfig{file=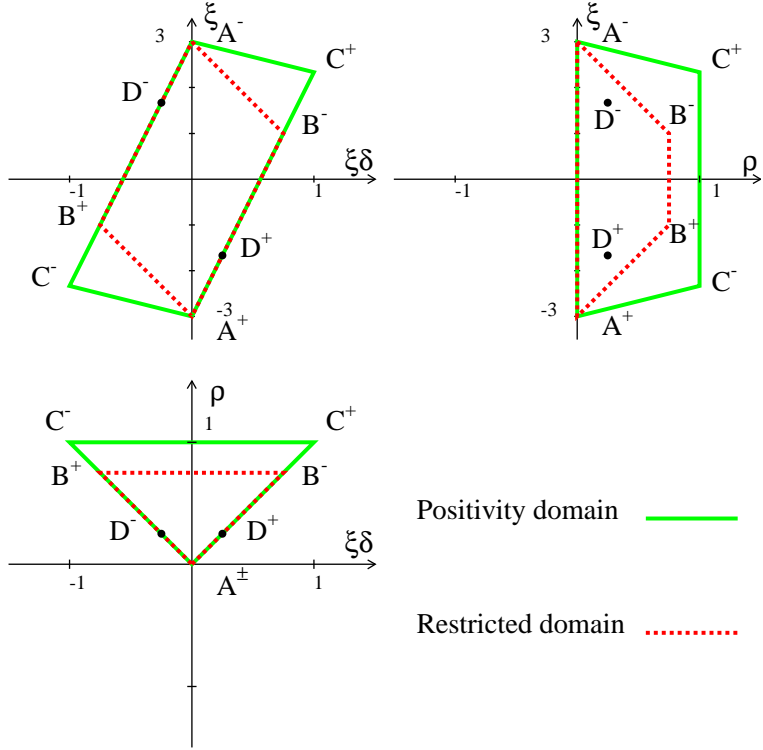,width=10cm}}
\caption{\protect\small The allowed domain in the space of the three parameters,
  $\rho$, $\xi$, and $\xi\delta$\label{fig:posdom}}
\end{figure}

A convenient way to express the relation between the Michel parameters and the  $g^{\gamma}_{ij}$
coupling constants is  to introduce~\cite{rouge95}
the six positive parameters
\begin{eqnarray}\nonumber
\alpha^+&=&|g^V_{RL}|^2+|g^S_{RL}+6g^T_{RL}|^2/16~,\\
\alpha^-&=&|g^V_{LR}|^2+|g^S_{LR}+6g^T_{LR}|^2/16~,\\\nonumber
\beta^+&=&|g^V_{RR}|^2+|g^S_{RR}|^2/4~,\\
\beta^-&=&|g^V_{LL}|^2+|g^S_{LL}|^2/4~,\\\nonumber
\gamma^+&=&(3/16)|g^S_{RL}-2g^T_{RL}|^2~,\\
\gamma^-&=&(3/16)|g^S_{LR}-2g^T_{LR}|^2~.\label{eq:gammadef}
\end{eqnarray}

They satisfy the relation
\begin{eqnarray}1&=&
\alpha^++\alpha^-+\beta^++\beta^-+\gamma^++\gamma^-\\\nonumber
&=&|g^V_{RR}|^2+|g^V_{LR}|^2+|g^V_{RL}|^2+|g^V_{LL}|^2\\\nonumber
&~&+\frac{1}{4}\,(|g^S_{RR}|^2+|g^S_{LR}|^2+|g^S_{RL}|^2+|g^S_{SS}|^2)
+3\,(|g^T_{LR}|^2+|g^T_{RL}|^2)~,
\end{eqnarray}
which means that the normalization is absorbed in the definition of
$G_{\tau l}$.

The $\rho$, $\xi$, and $\xi\delta$ parameters are given by
\begin{eqnarray}\nonumber\label{eq:micheldef}
\rho&=&\frac{3}{4}(\beta^++\beta^-)+(\gamma^++\gamma^-)~,\\
\xi&=&3\,(\alpha^--\alpha^+)+(\beta^--\beta^+)+\frac{7}{3}(\gamma^+-\gamma^-)~,\\\nonumber
\xi\delta&=&\frac{3}{4}(\beta^--\beta^+)+(\gamma^+-\gamma^-)~.
\end{eqnarray}

In geometrical terms, the point of coordinates $\rho$, $\xi$, and
$\xi\delta$, in the space of the parameters,
is the barycentre of six points, $A^\pm$, $B^\pm$, and $C^\pm$ with the
weights $\alpha^\pm$, $\beta^\pm$, and $\gamma^\pm$ respectively.
Since the point $B^-$ lies on the $A^+C^+$ segment, and $B^+$ on
$A^-C^-$, the allowed domain is just the tetrahedron $A^+A^-C^+C^-$ 
(Fig.~\ref{fig:posdom}).\\

\begin{table}[ht]
\caption{\label{table:gmax}\protect\small 
 Values of the coupling constants  in the Standard Model and
upper values compatible with the standard model prediction: $\xi=1$, $\rho=\xi\delta=3/4$}
\begin{center}
\begin{tabular}{ccccc} \hline
\rule{0em}{2.25ex}\rule[-1.5ex]{0em}{2.ex}&
$|g^V_{LL}|$&$|g_{LR}^V|$&$|g^V_{RL}|$&$|g^V_{RR}|$\\\hline
\rule{0em}{2.25ex}\rule[-1.1ex]{0em}{3.ex}maximum&$1$&$1$&$1$&$1$\\
\rule{0em}{2.25ex}\rule[-1.1ex]{0em}{3.ex}SM &$1$&$0$&$0$&$0$\\
\rule{0em}{2.25ex}\rule[-2.25ex]{0em}{3.ex}$\begin{array}{c}\hbox{maximum}\\
\rho=\xi\delta=3/4,\,\xi=1\end{array}$&
$1$&$0$&$1/2$&$0$\\\hline\hline
\rule{0em}{2.25ex}\rule[-1.5ex]{0em}{2.ex}&$|g_{LL}^S|$&$|g_{LR}^S|$&$|g_{RL}^S|$&$|g^S_{RR}|$\\\hline
\rule{0em}{2.25ex}\rule[-1.1ex]{0em}{3.ex}maximum&
$2$&$2$&$2$&$2$\\
\rule{0em}{2.25ex}\rule[-1.1ex]{0em}{3.ex}SM &$0$&$0$&$0$&$0$\\
\rule{0em}{2.25ex}\rule[-2.25ex]{0em}{3.ex}$\begin{array}{c}\hbox{maximum}\\
\rho=\xi\delta=3/4,\,\xi=1\end{array}$&
$2$&$0$&$2$&$0$\\\hline\hline
\rule{0em}{2.25ex}\rule[-1.5ex]{0em}{2.ex}&-&$|g^T_{LR}|$&$|g_{RL}^T|$&-\\\hline
\rule{0em}{2.25ex}\rule[-1.1ex]{0em}{3.ex}maximum&-&
$1/\sqrt{3}$&$1/\sqrt{3}$&-\\
\rule{0em}{2.25ex}\rule[-1.1ex]{0em}{3.ex}SM &-&$0$&$0$&-\\
\rule{0em}{2.25ex}\rule[-2.25ex]{0em}{3.ex}$\begin{array}{c}\hbox{maximum}\\
\rho=\xi\delta=3/4,\,\xi=1\end{array}$&-&
$0$&$1/2$&-\\\hline
 \end{tabular}
\end{center}
\end{table}

The Standard Model prediction, $|g^V_{LL}|=1$, implies  $\rho=\xi\delta=3/4\,,\,\xi=1$
and is represented geometrically by the
point $B^-$,  but $\beta^-=1$ does not imply
$|g^V_{LL}|=1$, and the  location of $B^-$ on the $A^+C^+$ segment
introduces further ambiguities.

The value $\beta^-=1$ is  also obtained if $|g^S_{LL}|=2$ and 
all the other constants equal to zero, and there is another
possibility to reproduce the 
parameters predicted by the Standard Model, namely 
 $\beta^-=\beta^+=\alpha^-=\gamma^-=0$, 
$\gamma^+=3/4$, and $\alpha^+=1/4$. So, with $\xi=1$, $\rho=\xi\delta=3/4$,
  $|g^V_{RL}|$ can reach an upper value of 1/2 and it is also possible that all the constants
vanish but $g^S_{RL}$ and $g^T_{RL}$. In this last case, we get
\begin{equation}
|g^S_{RL}|^2+12\,|g^T_{RL}|^2=4~,\hspace{1em}2\,|g^T_{RL}|=-|g^S_{RL}|\cos\phi_{ST} ~,  
\end{equation}
where $\phi_{ST}$ is the relative phase of the two amplitudes.
Accordingly, the upper possible values of  $|g^S_{RL}|$ and
$|g^T_{RL}|$ are 2 and 1/2 respectively.

The upper values of the constants, compatible with the Standard Model
prediction, are given in
Table~\ref{table:gmax}.
More stringent experimental limits~\cite{resdel} are  not the
consequences of the data but of additional hypotheses or constraints.
As the physically interesting region is the neighbourhood of $|g^V_{LL}|=1$,
no relevant bound on the non-standard $\tau$ left-handed
couplings can be 
extracted from the measurement of $\rho$, $\xi$ and $\xi\delta$
without additional hypotheses.   \\

Since the Standard Model prediction is represented by $B^-$ which is
located on $A^+C^+$, a convenient set of parameters is given by two
 equations of the $A^+C^+$ line and a third variable which specify the
 position on the segment. Using the equations of the faces $A^+C^+A^-$ and
$A^+C^+C^-$ gives  expressions which are positive for points inside
the tetrahedron. They are more usefully combined into
\begin{eqnarray}\label{eq:pdef}
{\cal P}_R^{\tau}&=&
\frac{1}{2}\,[\,1+\frac{\xi}{3}-\frac{16}{9}\,\xi\delta\,]
=\beta^++\alpha^-+\gamma^-\\\nonumber
&=&|g^V_{RR}|^2+|g^V_{LR}|^2+\frac{1}{4}\,|g^S_{RR}|^2+\frac{1}{4}\,|g^S_{LR}|^2+3|g^T_{LR}|^2~,\\
\label{eq:sdef}
{\cal S }_R^{\tau}&=&
\frac{2}{3}\,[
\rho-\xi\delta]=\beta^++\frac{4}{3}\,\gamma^-\\\nonumber
&=&|g^V_{RR}|^2+\frac{1}{4}\,|g_{RR}^S|^2
+\frac{1}{4}\,|g^S_{LR}-2g^T_{LR}|^2~,
\end{eqnarray}
which, owing to their normalization, can be interpreted as
 fractional contributions of  
$\tau$ right-handed couplings to its leptonic partial width and used 
to bound the $g^\gamma_{iR}$ amplitudes. For the third parameter,
$\xi$ or $\rho$ can be chosen.

\begin{figure}[ht]
\centerline{\epsfig{file=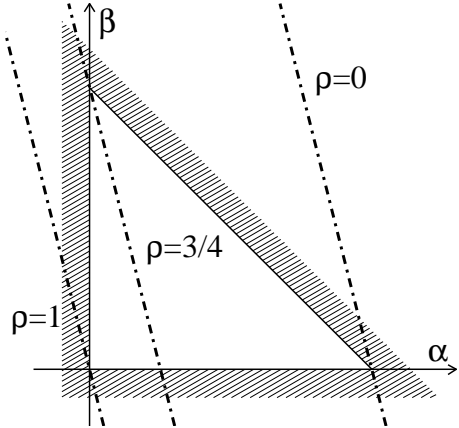,width=6cm}}
\caption{\protect\small The allowed domain in the $\alpha$, $\beta$ plane
 for a given value of $\rho$\label{fig:etapos}}
\end{figure}

The last Michel parameter $\eta$ can be written
\begin{eqnarray}\nonumber
\eta&=&2\,{\mathrm{Re}}\,[\,g^{V*}_{RL}\,(g^S_{LR}+6g^T_{LR})/4\,]
+2\,{\mathrm{Re}}\,[\,g^{V*}_{LR}\,(g^S_{RL}+6g^T_{RL})/4\,]\\
&~&+\,{\mathrm{Re}}\,[\,g^{V*}_{LL}\,g^S_{RR}/2\,]
+{\mathrm{Re}}\,[\,g^{V*}_{RR}\,g^S_{LL}/2\,]~.
\end{eqnarray}
This expression implies the inequality
\begin{equation}
|\eta|\leq\alpha+\frac{\beta}{2}\leq
1-\rho+\frac{1}{4}\beta_{\mathrm{max}}(\rho)~,
\end{equation}
where, $\alpha=\alpha^++\alpha^-$, $\beta=\beta^++\beta^-$, and 
 $\beta_{\mathrm{max}}(\rho)$ is the greatest value of
$\beta$, compatible with the value of
$\rho$. Fig.~\ref{fig:etapos} shows that $\beta_{\mathrm{max}}$ is
reached for $\gamma=1-\alpha-\beta=0$ when $\rho <3/4$ and for $\alpha=0$ for
$\rho>3/4$, leading to the rather weak bound:
\begin{equation}
|\eta|\leq 1-\frac{2}{3}\rho~~~(\rho\leq\frac{3}{4})~,\hspace{1em}|\eta|\leq
2(1-\rho)~~~(\rho\ge\frac{3}{4})
\end{equation}

\section{The restricted domain}
If only charged vector currents are present (\mbox{$g^S_{ij}=0,
g^T_{ij}=0$}), we have $\gamma^+=\gamma^-=0$ and the allowed domain is
the $A^+A^-B^+B^-$ tetrahedron that we call the restricted domain.

It is interesting to remark that, for a very wide class of models and
hypotheses, the allowed domain for the Michel parameters is this
restricted domain\footnote{
This result is almost obvious from the ``charge retention'' formalism,
where $\gamma^+$ and $\gamma^-$ are related to (neutral) tensor
currents~\cite{sirlin}, and Table~\ref{tab:neutral}.
} and that getting a point outside it requires the
conspiracy of scalar and tensor couplings~\cite{rouge95}.

We give below a list of general hypotheses leading to the restricted
domain. We use for that the Lorentz covariance of the charged and neutral currents
and the properties of the Fierz transformation only. More specific
constraints associated with the hypothesis of a single boson exchange
can be found in~\cite{musch} and~\cite{pich}.

When the allowed domain of the parameters is the restricted one, the
point $B^-$, which is the Standard Model prediction, is one of
its vertices, therefore there is a third positive quantity whose
vanishing is associated with this point.  We  use for it  $1-4\rho/3$.
Its precise physical meaning and the ambiguities in the interpretation
of the measurements depend nevertheless
on the hypothesis that lead to the restriction.\\

\noindent{\boldmath$g^S_{ij}=0$}\\

If all the scalar coupling constants vanish, the condition $\alpha^+=0$ implies $\gamma^+=0$,
therefore the complete domain is not allowed. Defining the new
parametrization
\begin{equation}
\alpha^+_{\mathrm{NS}}=|g^V_{RL}|^2~,\hspace{2em}
\gamma^+_{\mathrm{NS}}=3|g^T_{RL}|^2~,\hspace{2em}
\alpha^-_{\mathrm{NS}}=|g^V_{LR}|^2~,\hspace{2em}
\gamma^-_{\mathrm{NS}}=3|g^T_{LR}|^2~,
\end{equation}
we get the relations
\begin{eqnarray}
1&=&\alpha^+_{\mathrm{NS}}+\alpha^-_{\mathrm{NS}}+\beta^++\beta^-
+\gamma^+_{\mathrm{NS}}+\gamma^-_{\mathrm{NS}}~,\\[2ex]
\rho&=&\frac{3}{4}\,(\beta^++\beta^-)\nonumber
+\frac{1}{4}\,(\gamma^+_{\mathrm{NS}}+\gamma^-_{\mathrm{NS}})~,\\
\xi&=&3\,(\alpha^-_{\mathrm{NS}}-\alpha^+_{\mathrm{NS}})
+(\beta^--\beta^+)-\frac{5}{3}\,(\gamma^+_{\mathrm{NS}}-\gamma^-_{\mathrm{NS}})~,\\\nonumber
\xi\delta&=&\frac{3}{4}\,(\beta^--\beta^+)
+\frac{1}{4}\,(\gamma^+_{\mathrm{NS}}-\gamma^-_{\mathrm{NS}})~.
\end{eqnarray}
The point of coordinates $\rho$, $\xi$, and $\xi\delta$ is now the
barycentre of $A^\pm$, $B^\pm$, and two new points
$C^\pm_{\mathrm{NS}}\equiv D^\pm$ whose
coordinates are ($1/4$, $\mp 5/3$, $\pm 1/4$) but, since $D^+$ is located
on the $A^+B^-$ segment and $D^-$ on $A^-B^+$ (Fig.~\ref{fig:posdom}), the allowed region
is  the restricted domain. The third constraint can be written,
in terms of the $g^\gamma_{ij}$'s, as
\begin{equation}
1-\frac{4}{3}\,\rho=|g^V_{RL}|^2+|g^V_{LR}|^2 +2\,(|g^T_{RL}|^2+|g^T_{LR}|^2)~.
\end{equation}
Measurements represented by  $B^-$ imply a V-A structure.
The combination of charged, vector and tensor currents belongs to
this class.\\
 
\noindent{\boldmath$g^T_{ij}=0$}\\

If $g^T_{ij}=0$, we define
\begin{equation}
\alpha^+_{\mathrm{NT}}=|g^V_{RL}|^2~,\hspace{2em}
\gamma^+_{\mathrm{NT}}=\frac{1}{4}\,|g^S_{RL}|^2~,\hspace{2em}
\alpha^-_{\mathrm{NT}}=|g^V_{LR}|^2~,\hspace{2em}
\gamma^-_{\mathrm{NT}}=\frac{1}{4}\,|g^S_{LR}|^2~,
\end{equation}
and obtain
\begin{eqnarray}
1&=&\alpha^+_{\mathrm{NT}}+\alpha^-_{\mathrm{NT}}+\beta^++\beta^-
+\gamma^+_{\mathrm{NT}}+\gamma^-_{\mathrm{NT}}~,\\[2ex]\nonumber
\rho&=&\frac{3}{4}\,(\beta^++\beta^-
+\gamma^+_{\mathrm{NT}}+\gamma^-_{\mathrm{NT}})~,\\
\xi&=&3\,(\alpha^-_{\mathrm{NT}}-\alpha^+_{\mathrm{NT}})
+(\beta^--\beta^+)+(\gamma^+_{\mathrm{NT}}-\gamma^-_{\mathrm{NT}})~,\\\nonumber
\xi\delta&=&\frac{3}{4}\,(\beta^--\beta^++
\gamma^+_{\mathrm{NT}}-\gamma^-_{\mathrm{NT}})~.
\end{eqnarray}
Here the points $C^+_{\mathrm{NT}}$ and $C^-_{\mathrm{NT}}$ coincide with the points $B^-$ and $B^+$
respectively. The allowed domain is then the restricted one. The third
constraint is
\begin{equation}
1-\frac{4}{3}\,\rho=|g^V_{RL}|^2+|g^V_{LR}|^2~.
\end{equation}

The vector couplings only are bounded 
by the measurement of $\rho$, due to the 
ambiguity between $B^\pm$ and $C^\mp_{\mathrm{NT}}$. 

The combination of charged, vector and scalar currents and neutral,
vector currents belongs to this class.\\

\noindent{\bf \boldmath $g^S_{RL}=2g^T_{RL}$, $g^S_{LR}=2g^T_{LR}$}\\

The restriction of the domain is evident from Eq.~\ref{eq:gammadef}.
The third constraint can be written 
\begin{equation}
1-\frac{4}{3}\,\rho\,=\,|g^V_{RL}|^2+|g^V_{LR}|^2+\frac{1}{4}\,(|g^S_{LR}|^2+|g^S_{RL}|^2)
+3\,(|g^T_{LR}|^2+|g^T_{RL}|^2)~.
\end{equation} 
Measurements represented by  $B^-$ imply a V-A structure.

The combination of charged, vector currents and neutral, vector and
scalar currents belongs to this class.\\

\noindent{\bf ``V-A plus something''}\footnote{The title of this
  paragraph is borrowed from C.~Nelson~\cite{nelson}.}\\

In an especially interesting family of models~\cite{pich} the decay is
described by the addition of a
single, non-standard contribution to the Standard Model amplitude. 
Their predictions are perfectly transparent in the present geometrical
presentation.
They follow at once from the definitions of the parameters and the properties
displayed in Table~\ref{tab:neutral}.

-If the non-standard contribution is a charged vector current,
all the restricted domain is allowed. The only new prediction is
$\eta=0$.

-The hypothesis of an additional charged scalar current belongs to the second of
the above defined classes with the further conditions
$\alpha^\pm_{\mathrm{NT}}=0$. The allowed domain is then the $B^+B^-$ segment
 ($\rho=\delta=3/4$), since the points
$C^\pm_{\mathrm{NT}}$ and $B^\mp$ are identical.

-The contribution of a neutral vector is included in the same class
with $\alpha^\pm_{\mathrm{NT}}=\gamma^\pm_{\mathrm{NT}}=0$. The
allowed domain is again the  $B^+B^-$ segment.

-The hypothesis of an additional neutral scalar leads to $\gamma^+=\gamma^-=0$ (third
class) and $\beta^+=0$. The allowed domain is the two-dimensional
 one, spanned by the points $A^+$, $A^-$, and $B^-$.
The corresponding condition is $\rho=\xi\delta$ and the bound on
$\eta$ is  stricter: $|\eta|\le 1-4\rho/3$.

\section{Looking for new physics}\label{sec:lnp}
In the standard approach described in section~\ref{cha:intro}, the  indicators of new
physics constructed with the Michel parameters are, besides  the $\eta$ parameter
itself, the two positive quantities ${\cal P}^\tau_R$ and  ${\cal
  S}^\tau_R$, defined by Eqs.~\ref{eq:pdef} and~\ref{eq:sdef}, which
bound the coupling constants and
can be interpreted  as (non-independent) contributions of new physics
to the $\tau$ decay.

Using the world-average values of the parameters~\cite{pdg}, under the hypothesis of $e$-$\mu$
universality, yields
\[ {\cal P}^\tau_R=0.006\pm 0.028~,\hspace{3em}{\cal
  S}^\tau_R=0.001\pm 0.016~.\] 

If hypotheses are made that reduce the dimensionality of the domain,
the determination of the parameters can be improved by a constrained fit. For instance, if
the domain is the $B^+B^-$ segment, there is only one free parameter.
Neglecting the correlations between the measurements and taking
advantage of the near equality of the errors on $\xi$ and
$4\xi\delta/3$, this parameter is merely their average $\bar\xi$
and the quantity ${\cal P}^\tau_R$ reduces to
\begin{equation}
\label{eq:xibardef} 
\frac{1}{2}\,[\,1-\bar\xi\,]=\frac{1}{2}\,[\,1-\frac{\xi}{2}-\frac{2}{3}\xi\delta\,]~.
\end{equation}

It is noteworthy that the same strategy can be followed under the much
weaker hypotheses that imply the restricted domain.
This is due to the fact that the measured value of $\rho$ forces
the  point which represents the measurements in the parameter space to lie on the edge of
the domain.

Quantitatively, the difference of $\xi$ and $4\xi\delta/3$ is bounded by
the inequality
\begin{equation}
\big |\xi-\frac{4}{3}\,\xi\delta\big |\leq 3\big
  (1-\frac{4}{3}\,\rho\big )~.
\end{equation}
From the values of the errors~\cite{pdg} $\sigma(\xi-4\xi\delta/3)=
0.044$ and $4\sigma(\rho)=0.036$, it is clear that there is no
physically relevant information in $\xi-4\xi\delta/3$ and that even the $\tau$-$l$
universality prediction~\cite{musch} $\delta=3/4$  is better tested    
by the measurement of $\rho$ than by comparing $\xi$ and $\xi\delta$.

Assuming only the restricted domain, the indicators of new physics are
\[
1-\frac{4}{3}\,\rho=0.004\pm 0.012~,\hspace{2em}
\frac{1}{2}\,[\,1-\bar\xi\,]=0.002\pm 0.011~.  \]
They bring a clear improvement of the sensitivity with respect to  ${\cal
  P}^\tau_R$ and   ${\cal S}^\tau_R$.

It must be noted that $1-\bar\xi$ is not strictly positive but that
excursions of $\bar\xi$ beyond 1 are severely limited by the
inequality
\begin{equation}
\bar\xi-1\leq\frac{1}{2}\,\big (1-\frac{4}{3}\,\rho\big)~,
\end{equation}
since $\sigma(\bar\xi )=0.022$ and $2\sigma(\rho)/3=0.006$.
\section{Conclusion}
A complete study of the constraints on the Michel parameters and the
ambiguities of their interpretation has been presented.

It has been shown that, for a very wide class of hypotheses and
models, which cause the same restriction of the parameter domain,
 the best indicators of new physics are the combinations,
$1-4\rho/3$ and
$(1-\bar\xi)/2=(1/2-\xi/4-\xi\delta/3)$.

Compared to the customary estimators, their sensitivities are roughly
twice better. The third available parameter, $\xi-4\xi\delta/3$, is better determined by
the geometry of the domain and the value of $\rho$ than by its measurement. 

\appendix
\section{Appendix: Measurement of the parameters}
In the numerical exercise above, the error correlations were neglected
and the $e$-$\mu$ universality was assumed. We will discuss briefly
this two approximations.

At low energy, where the $\tau$'s are unpolarized, the $\rho$ and
$\eta$ parameters are determined by the single-lepton laboratory
energy distributions and the $\xi$ and $\xi\delta$ parameters by the spin-correlated
$\tau^+$ $\tau^-$ decay distributions. Estimates of the covariance
matrices for measurements at 4~GeV and 10~GeV can be found in~\cite{fetschb}.
 
At the Z~peak, the $\tau$ polarization makes ambiguous the
interpretation of a single tau leptonic-decay distribution and,
since the transverse spin correlations depend on the Z couplings, only
the helicity correlation of $\tau^+$ and $\tau^-$, which is equal to
-1, is used. If the decay distribution of a $\tau^-$ in the channel
$a$, is written,
\begin{equation}
W_a(x^-)=f_a(x^-)+P_\tau\,g_a(x^-)~,
\end{equation}
the correlated distribution for the $a$ and $b$ channels reads then
\begin{equation}
W_{ab}(x^-,x^+)=f_a(x^-)f_b(x^+)+g_a(x^-)g_b(x^+)
+\,P_\tau\,[\,f_a(x^-)g_b(x^+)+f_b(x^+)g_a(x^-)\,]~,
\end{equation}
where $P_\tau$ is the $\tau^-$ polarization.

For a hadronic decay, with the notations of~\cite{omega},
 $x^\pm$ is the optimal variable
$\omega$ and the decay distribution is
\begin{equation}
W(\omega)=\hat{f}(\omega)[1+\xi_h\,P_\tau\,\omega]~,
\end{equation}
where the $\xi_h$ parameter  is equal to 1 in the standard
model\footnote{
Since $P_{\tau^+}=-P_{\tau^-}$ and $\xi_h^{\tau^+}=-\xi_h^{\tau^-}$,
the decay distribution is independent of the $\tau$ charge. In the
case of $\tau\to\nu 3\pi$, this last property is true only if
no pseudoscalar variable constructed from the $\pi$ momenta is used in
the definition of $\omega$. A more general analysis is presented
in~\cite{argus}.}.

For a leptonic decay, $x^\pm$ is the normalized energy of the charged
lepton, $y=(E_l/E_l^{\mathrm{max}})_{\mathrm{LAB}}$. 
Defining the parameters,
\begin{equation}
\tilde\rho=1-\frac{4}{3}\rho\,,\hspace{3em}\tilde\delta=\xi-\frac{4}{3}\xi\delta\,,\hspace{3em}
\tilde\eta=\frac{m_l}{m_\tau}\eta\,,
\end{equation}
which vanish in the Standard model, and the functions
\begin{eqnarray}&~&
h_0(y)=\frac{1}{3}(5-9y^2+4y^3)\,,\hspace{6em}
h_1(y)=\frac{1}{3}(1-9y^2+8y^3)\,,
\\&~&
h_2(y)=\frac{1}{3}(1-12y+27y^2-16y^3)\,,\hspace{5em}
h_3(y)=12(1-y)^2\,,
\end{eqnarray}
the decay distribution reads
\begin{equation}\label{eq:declab}
W(y)
=f_{\rho,\eta}(y)+P_\tau
g_{\xi,\xi\delta}(y)
=\frac{1}{1+4\tilde\eta}\big\{h_0(y)
+[\tilde\rho+ P_\tau\xi]h_1(y)+
\tilde\delta P_\tau h_2(y)+\tilde\eta h_3(y)
\big \}~.
\end{equation}
The presence of the polarization allows the measurement of a new
parameter, $\tilde\delta P_\tau$, but, as previously mentioned, it also
introduces an ambiguity in the interpretation of the first one which
is now $\tilde\rho+P_\tau\xi$.

The same kind of ambiguity also arises in the $e$-$\mu$ correlation.
For $P_\tau=0$, only the parameters, $\delta_\mu$,
$\delta_e$,
and the product   $\xi_\mu\xi_e$
 are measurable~\cite{fetschb}. For  $P_\tau\neq 0$,
neglecting $\tilde\eta$ for the sake of clarity, and keeping only
terms of the first order in the Standard Model violating parameters,
$\tilde\rho$, $\tilde\delta$ and $1-\xi$, the correlated distribution
can be written
\begin{eqnarray}\label{eq:decemu}
\lefteqn{W(y_e,y_\mu)
\sim h_0(y_\mu)h_0(y_e)-h_1(y_\mu)h_1(y_e)
+[\tilde\rho_e+P_\tau\xi_e]h_0(y_\mu)h_1(y_e)}\hspace{4em}\\\nonumber
&&+[\tilde\rho_\mu+P_\tau\xi_\mu]h_0(y_e)h_1(y_\mu)
+[\xi_\mu+\xi_e+P_\tau(\tilde\rho_e+\tilde\rho_\mu)]h_1(y_\mu)h_1(y_e)\\\nonumber&&
+\tilde\delta_e h_2(y_e)[h_1(y_\mu)+P_\tau h_0(y_\mu)]
+\tilde\delta_\mu h_2(y_\mu)[h_1(y_e)+P_\tau h_0(y_e)]~.
\end{eqnarray}
Even if $P_\tau$ is known, there is only three measurements to
determine the four parameters $\rho_\mu$, $\rho_e$, $\xi_\mu$ and
$\xi_e$. The ambiguity is displaced but not
suppressed\footnote{Therefore, the analysis~\cite{resopal} necessarily
  uses additional hypotheses.}.

Using several known values of the polarization, all the parameters can
be determined from single-decay distributions. At the Z peak, and/or
with polarized beam, the $\tau$ polarization is a function of the
production angle $\theta$, hence the Michel parameters can be measured
by the $\theta$-$y$ correlation~\cite{ressld}.

The other measurements~[15,22-24]
 use all the hadron-hadron, lepton-lepton and
hadron-lepton final states to obtain the complete set of parameters up
to a global sign ambiguity which is solved, for instance, by the
result of~\cite{ressld}.

To calculate the covariance matrix $V$ of the Michel parameter
measurements,   we assume that,
in an ideal experiment at the Z
peak, similar to~[15,22-24],
 all the decays into $\mu$, $e$, $\pi$, $\rho$ and
$a_1$ and their correlations are used to determine the Michel
parameters, the hadronic, $\xi_h$ parameters and the $\tau$
polarization. Asymptotically $V$ is given by
\begin{equation}
\left (V^{-1}\right )_{ij} \sim -\sum_s\, N_s\!\int\!\!\int
W_s\frac{\partial^2\log W_s}{\partial\alpha_i\partial\alpha_j}dx^+dx^-~,
\end{equation}
where $s$ labels  one of the twenty classes of events, $e$-$\mu$, $e$-$e$,
$e$-$\pi$, etc, and $N_s$ is the number of events in the class.
The computation is made straightforward by the fact that
 the distributions are quadratic
functions of all the parameters, except $\eta$
which appears in the normalizations. 
\begin{table}[h]
\caption{\protect\small Ideal statistical errors on the Michel
  parameters (in \%)  and their correlation 
coefficients for 2$\times
  10^5$ $\tau^+\tau^-$ pairs at the Z peak\label{tab:covid}}
\begin{center}
\normalsize
\begin{tabular}{@{}l@{}c@{}c@{}c@{}c@{}}
\rule{2em}{0ex}&\rule{4.44em}{0ex}&\rule{4.44em}{0ex}&\rule{4.44em}{0ex}&\rule{4.44em}{0ex}\\
\hline
\rule{0em}{2.75ex}\rule[-1.5ex]{0em}{2.ex}&
$\sigma(\rho)$&$\sigma(\xi)$&$\sigma(\xi\delta)$&$\sigma(\xi-\frac{\textstyle 4}{\textstyle 3}\xi\delta)$
\\\hline
\rule{0em}{2.5ex}\rule[-2.ex]{0em}{1.5ex}
$\mu$&2.3&6.0&4.0&7.3\\
\rule{0em}{1.75ex}\rule[-1.ex]{0em}{2.ex}
$e$&1.3&5.5&3.9&7.2
\\\hline
\end{tabular}
\begin{tabular}{@{}l@{}c@{}c@{}c@{}}
\rule{2em}{0ex}&\rule{5.92em}{0ex}&\rule{5.92em}{0ex}&\rule{5.92em}{0ex}\\
\hline
\rule{0em}{2.75ex}\rule[-1.5ex]{0em}{2.ex}&
$C(\rho,\xi)$&$C(\rho,\xi\delta)$&$C(\xi,\xi\delta)$\\\hline
\rule{0em}{2.5ex}\rule[-2.ex]{0em}{1.5ex}
$\mu$&0.24&0.15&0.18\\
\rule{0em}{1.75ex}\rule[-1.ex]{0em}{2.ex}
$e$&-0.21&-0.06&0.10\\\hline
\end{tabular}
\end{center}
\end{table}

The computed covariance matrix is  similar to its
estimations~\cite{fetschb} for measurements at lower energy.
The  largest correlation coefficients are
$C(\rho_\mu,\eta_\mu)=0.82$ and $C(\xi_\mu,\eta_\mu)=0.42$.
The numerical values relevant for the analysis of
Section~\ref{sec:lnp} are given in Table~\ref{tab:covid}.

For the $\tau\to e\nu\bar\nu$ channel, the inequality
$\sigma(\tilde\delta)>3\sigma(\tilde\rho)$ is satisfied and the
weights of $\xi$ and $4\xi\delta/3$ in their optimal combination are
 0.47 and 0.53 respectively. All the hypotheses made in
Section~\ref{sec:lnp} are verified.

For the $\tau\to\mu\nu\bar\nu$ channel, the weights are 0.43 and 0.57
but, owing to the correlations with $\eta_\mu$, the error on
$\tilde\delta$ is slightly smaller than $3\sigma(\tilde\rho)$.
However, in a more realistic estimation, the inefficiencies
in the identification of the various decay channels reduce the
statistics for the classes of events with two analysed decays
and increase it for the
 events with only one identified decay which contribute mainly to the
 measurement of $\rho$.
Therefore, the same analysis scheme remains basically valid.

From an experimental point of view, the universality hypothesis allows
to constrain the value of $\eta_\mu$  by the measurement of the
$\tau\to e\nu\bar\nu$ and $\tau\to\mu\nu\bar\nu$ branching
ratios~\cite{resdel,resopal}.

If the variation of the parameters is limited to the above defined
restricted domain, the deviations of the parameters from their
Standard Model values must have the same sign in the $e$ and $\mu$
channels. Therefore the universality hypothesis can perhaps reduce
the sensitivity to these deviations but complete  cancelations
are not  possible.

\end{document}